\documentclass[twocolumn,preprintnumbers,amsmath,amssymb,showpacs]{revtex4}
\usepackage{epsfig}
\usepackage{color}
\begin{document}
\setlength{\voffset}{1.0cm}
\title{Kink dynamics, sinh-Gordon solitons and strings in AdS$_3$ \\ from the Gross-Neveu model}
\author{Andreas Klotzek\footnote{andreas@theorie3.physik.uni-erlangen.de}}
\author{Michael Thies\footnote{thies@theorie3.physik.uni-erlangen.de}}
\affiliation{Institut f\"ur Theoretische Physik III,
Universit\"at Erlangen-N\"urnberg, D-91058 Erlangen, Germany}
\date{\today}
\begin{abstract}
Guided by a study of kink-antikink scattering in the Gross-Neveu model and other known solutions of the Hartree-Fock approach of 
a particularly simple type, we demonstrate a quantitative relationship between three different problems: Quantized 1+1-dimensional 
fermions in the large $N$ limit, solitons of the classical sinh-Gordon equation and classical strings moving in 3-dimensional anti de Sitter
space. Aside from throwing light on the relationship between quantum field theory and classical physics, this points to the full solvability
of the dynamical $N$-kink-antikink problem in the Gross-Neveu model.
\end{abstract}
\maketitle
\section{Introduction}\label{sect1}
The simplest variant of the Gross-Neveu (GN) model family \cite{L1} consists of $N$ species of massless Dirac fermions in 1+1 dimensions, 
interacting via a scalar-scalar four-fermion interaction,
\begin{equation}
{\cal L} =  \sum_{k=1}^N \bar{\psi}_k i\partial \!\!\!/ \psi_k + \frac{g^2}{2} \left( \sum_{k=1}^N \bar{\psi}_k \psi_k \right)^2.
\label{I1}
\end{equation}
This model has asymptotic freedom, no scale, and a discrete chiral symmetry $\psi \to \gamma_5 \psi$ which gets broken spontaneously in 
the vacuum, yielding a dynamical fermion mass $m$ via dimensional transmutation. Throughout this paper we will be exclusively dealing with 
the 't~Hooft limit $N\to \infty, Ng^2 =$ const. \cite{L2}. As is well known, the attractive interaction gives rise to a marginally bound scalar
fermion-antifermion state (the $\sigma$-meson) with mass 2$m$ and to a rich variety of multi-fermion bound states (baryons) \cite{L3,L4}. 
Moreover, the model features a non-trivial phase diagram as a function of temperature and chemical potential with three distinct phases 
(massless and massive Fermi gas, baryon crystal) meeting at a tricritical point \cite{L5}. It is of interest not only as a toy model for strong 
interaction particle physics, but also because of its almost literal recurrence in condensed matter systems such as conducting polymers,
carbon nanotubes or quasi-one-dimensional superconductors \cite{L6}. In the large $N$ limit, baryons as well as baryonic matter and the
phase diagram can be determined with semiclassical methods, notably the relativistic Hartree-Fock (HF) method \cite{L7}. Relativity enters
in two ways --- use of the Dirac equation instead of the Schr\"odinger equation, and taking into account the filled, interacting Dirac sea. 
Recently we have started to address time-dependent questions by generalizing this approach to time-dependent Hartree-Fock (TDHF) 
\cite{L8}. In a first step, the boosted baryon was considered, demonstrating a covariant energy-momentum relation and deriving exact 
structure functions as fermion momentum distributions in the infinite momentum frame \cite{L9}. In the present work, we proceed to the next
level of complication and study baryon-baryon scattering, an issue which has not yet been discussed in any detail in the large $N$ limit.
This may well be the first concrete realization of Witten's vision about baryon-baryon interactions in the large $N$ limit, originally developed
in a non-relativistic context \cite{L9a}. Aside from the genuine interest in the scattering problem of composite, relativistic bound states, we hope
to get further insights into the mathematical structure of the GN model by enlarging the spectrum of questions addressed in the framework of the
TDHF approach.

Before embarking on this problem, it may be worthwhile to recapitulate some of the experience gained with previous applications of the
HF approach. The basic mathematical problem can be formulated in a single line as
\begin{equation}
\left( i \partial \!\!\!/ - S\right) \psi_{\alpha} =0, \qquad S=-g^2 \sum_{\beta}^{\rm occ} \bar{\psi}_{\beta}\psi_{\beta},
\label{I2}
\end{equation}
where the $\psi_{\alpha}$ are $c$-number spinors (single particle wave functions).
Since the sum over occupied states includes the Dirac sea, this is an infinite system of coupled, non-linear partial differential equations.
At finite temperature and chemical potential, a similar formula applies, but the sum includes thermal occupation numbers.
Except in the case of a homogeneous mean field $S$ which acts like a dynamical mass, the solution of Eqs.~(\ref{I2}) is highly non-trivial.
Nevertheless, for the model defined in Eq.~(\ref{I1}), closed analytical solutions have been found in all cases studied so far. If one examines
in detail how self-consistency is achieved, one notices that all known solutions fall into two classes: Type I solutions are such that the 
contribution from every occupied single particle state is either zero or proportional to the full mean field $S$,
\begin{equation}
\bar{\psi}_{\alpha} \psi_{\alpha} = \lambda_{\alpha} S.
\label{I3}
\end{equation}
The self-consistency condition then becomes space-time independent,
\begin{equation}
- Ng^2 \sum_{\alpha}^{\rm occ} \lambda_{\alpha} = 1.
\label{I4}
\end{equation}
Aside from the trivial case of the vacuum, type I solutions have been found for the kink at rest \cite{L3,L10} or in flight \cite{L9} and the kink 
crystal at zero temperature \cite{L11}  relevant 
for the ground state at finite density. In type II solutions, the scalar condensate for each occupied state involves two different space-time 
functions. Without loss of generality, this can be written as
\begin{equation}
\bar{\psi}_{\alpha} \psi_{\alpha} = \lambda_{\alpha} S + \lambda_{\alpha}' S',
\label{I5}
\end{equation}
where $S$ is the self-consistent potential and $S'$ a second, different function. This yields two space-time independent self-consistency
conditions,
\begin{equation}
- Ng^2 \sum_{\alpha}^{\rm occ} \lambda_{\alpha} = 1, \quad \sum_{\alpha}^{\rm occ} \lambda_{\alpha}' = 0.
\label{I6}
\end{equation}
Known type II solutions are the kink-antikink baryons \cite{L3,L10}, the kink crystal at finite temperature \cite{L12} and the time-dependent
breather \cite{L3} to be discussed below in more detail. Incidentally, all HF solutions of the massive GN model (i.e., Lagrangian (\ref{I1})
supplemented by a term $-m_0 \sum_{k=1}^N \bar{\psi}_k \psi_k$) are also of type II \cite{L13,L14,L15}, and no
solution of type III or higher is known in the GN model which would require more than two functions and hence more than two space-time
independent
self-consistency conditions. As will become clear later on, this classification of HF solutions is useful if one wants to relate the quantum 
theory to classical GN models with a small number of flavors. 
 
In this paper, we shall focus on the dynamics of kinks, supplementing the known single kink and kink crystal solutions by kink-antikink
scattering and briefly commenting on the generalization to $N$ kinks and antikinks. It turns out that this restriction to kink dynamics 
is at the same time a restriction to 
type I HF solutions. For this class of particularly simple (though exact) solutions of the large $N$, massless GN model, we will identify an
effective bosonic theory whose soliton solutions are closely related to self-consistent HF potentials, using methods developed by Neveu and 
Papanicolaou in their proof of integrability of the classical one- and two-flavor GN models \cite{L16}. The relevant equation is the sinh-Gordon
equation, the hyperbolic version of the more familiar sine-Gordon equation. This in turn is the key for mapping type I solutions of the GN model
onto classical string theory in 3-dimensional anti de Sitter space (AdS$_3$), following recent work of Jevicki and collaborators
\cite{L17,L18,L19}. In 
this way we hope to show that in spite of the simplicity of the GN model, its mature age and the considerable amount of work devoted to it, 
there is still room for new insights and surprises.

This paper is organized as follows. In Sec.~\ref{sect2}, we solve the kink-antikink scattering problem with the help of the known breather.
Sec.~\ref{sect3} aims at identifying the sinh-Gordon equation as underlying effective bosonic theory. In Sec.~\ref{sect4}, we show how to map
solutions of the GN model onto strings moving in AdS$_3$ and illustrate this mapping with a few examples. In Sec.~\ref{sect5} we summarize
our findings and identify promising directions for future work.

\section{Kink-antikink scattering}\label{sect2}
Our starting point is the kink-antikink breather solution of the GN model discovered by Dashen, Hasslacher and Neveu (DHN) 
\cite{L3}. Since the inverse scattering method is not available for time-dependent semi-classical solutions, these authors 
had to guess the scalar potential by analogy with the well-known sine-Gordon breather. They then show that 
a self-consistent solution can be found whose quantized fluctuations yield the spectrum of kink-antikink type baryons.
At the end of their paper, DHN mention that the imaginary choice $\epsilon= i/v$ of a certain parameter governing the frequency of the
breather should describe scattering of a kink-antikink pair with velocities $\pm v$ in their center-of-mass (cm) frame. We take
up this suggestion here and analyze the scattering of a kink and an antikink in greater depth, using the language of the TDHF approach.
Before doing this however, let us review the results for the DHN breather which are relevant for the analytic continuation in $\epsilon$.
This is also necessary because the formulae given in \cite{L3} do not seem to be fully consistent, either due to typos or to some unspecified
conventions. 

The breather is a solution of the Dirac equation
\begin{equation}
\left( i \partial \!\!\!/ - S \right) \psi = 0
\label{1}
\end{equation}
and will be given in the representation
\begin{equation}
\gamma^0 = - \sigma_1, \quad \gamma^1 =  i\sigma_3, \quad \gamma_5 = \gamma^0 \gamma^1 = - \sigma_2
\label{2}
\end{equation}
of the $ \gamma$ matrices. Units in which the vacuum fermion mass is 1 will be used throughout this work. 
The scalar potential can be written as
\begin{equation}
S = 1 + \xi f_2 + \eta f_4
\label{3}
\end{equation}
with
\begin{eqnarray}
f_2 & = & f_4 \cos \Omega t ,
\nonumber \\
f_4 & = & (\cosh Kx + a \cos \Omega t + b)^{-1}.
\label{4}
\end{eqnarray}
The parameters appearing here are defined and related as follows,
\begin{eqnarray}
\Omega & = & \frac{2}{\sqrt{1+\epsilon^2}}, \quad K \ = \ \epsilon \Omega,
\nonumber \\
\xi & = &  -2a, \quad \eta \ = \ - \frac{1}{2} b K^2.
\label{5}
\end{eqnarray}
$a$ is the solution of the equation
\begin{equation}
0 = \eta^2+K^2 \left(1-b^2 \right) + \Omega^2 a^2,
\label{6}
\end{equation}
where we choose the positive square root. $S$ is fully specified if the parameters $\epsilon$ and $b$ are given.
The Dirac equation (\ref{1}) with potential (\ref{3}) has two types of solutions.
First, there is a continuum of wave-like solutions of the form
\begin{equation}
\psi_k = \left( \begin{array}{c} u_k \\ v_k \end{array} \right) e^{i(kx-\omega t)} 
\label{7}
\end{equation}
with $\omega = \pm \sqrt{k^2+1}$. Like in the single baryon problem, these scattering states have no 
reflected wave, showing that also the time-dependent potential is transparent.
DHN give the explicit form of $\psi_k$ in terms of $f_2,f_4$ from Eq.~(\ref{4}) and two further functions
\begin{equation}
f_1 = f_4 \sinh Kx, \quad f_3 = f_4 \sin \Omega t .
\label{8}
\end{equation}
We have redetermined the coefficients with the following results,
\begin{eqnarray}
u_k & = & {\cal N}_k \left( 1 + \frac{ iK}{2k} f_1 - \frac{ ia}{k} f_2 + \frac{a \Omega}{2k} \frac{1+ik}{\omega} f_3
+ \frac{ i\eta}{2k} f_4 \right),
\nonumber \\
v_k & = & {\cal N}_k  \left( - \frac{1+ik}{\omega} - \frac{K}{2k}\frac{i-k}{\omega} f_1- \frac{a}{k}\frac{i-k}{\omega}f_2 \right.
\nonumber \\
& &  \left. + \frac{a\Omega}{2k} f_3 + \frac{\eta}{2k}\frac{ i-k}{\omega} f_4 \right).
\label{9}
\end{eqnarray} 
They deviate from Eq.~(4.6) of Ref.~\cite{L3}, but one can easily check that Eqs.~(\ref{7},\ref{9}) do solve the Dirac equation. The second type of
solution are bound states. DHN find two distinct states which can be expressed in terms of the functions
\begin{eqnarray}
\varphi_1 & = & f_4 \cos \Omega t/2 \cosh Kx/2 ,
\nonumber \\
\varphi_2 & = & f_4 \sin \Omega t/2 \sinh Kx/2 ,
\nonumber \\
\varphi_3 & = & f_4 \sin \Omega t/2 \cosh Kx/2 ,
\nonumber \\
\varphi_4 & = & f_4 \cos \Omega t/2 \sinh Kx/2 .
\label{10}
\end{eqnarray}
Two orthogonal solutions of the Dirac equation are 
\begin{eqnarray}
\psi_0^{(1)} & = & {\cal N}_0 \left( \begin{array}{r} \varphi_1 + c_4 \varphi_4
 \\ c_2 \varphi_2 + c_3 \varphi_3
 \end{array} \right),
\nonumber \\
\psi_0^{(2)} & = & {\cal N}_0 \left( \begin{array}{r}  -c_2 \varphi_2 + c_3 \varphi_3
   \\  \varphi_1 - c_4 \varphi_4    \end{array} \right).
\label{11}
\end{eqnarray}
Here, our coefficients
\begin{eqnarray}
c_2 & = & - i\epsilon \frac{1-b-\eta/2}{a-\eta/2},
\nonumber \\
c_3 & = & \frac{ i\Omega}{2} \frac{1+b-a}{1+b+\eta/2},
\nonumber \\
c_4 & = & - \frac{K}{2} \frac{1-a-b}{a-\eta/2},
\label{12}
\end{eqnarray}
differ from those of DHN in the overall signs of $c_3$ and $c_4$. 
Self-consistency of this solution can be established as follows: DHN find that $\bar{\psi}_k\psi_k$ for each occupied continuum state yields 
two contributions, one proportional to $S$ and one proportional to $f_4$. The discrete states yield only a single contribution proportional
to $f_4$. By relating the parameter $b$ to the occupation fraction of the discrete states, one gets a self-consistent result where the
$f_4$ terms cancel. According to the classification of TDHF solutions given in the introduction, this shows that the breather is a type II
solution of the GN model. Only for the value $b=0$, the contributions proportional to $f_4$ would vanish and the breather would be 
a type I solution. However $b=0$ is ruled out since it is incompatible with a real potential $S$.

\begin{figure}
\begin{center}
\epsfig{file=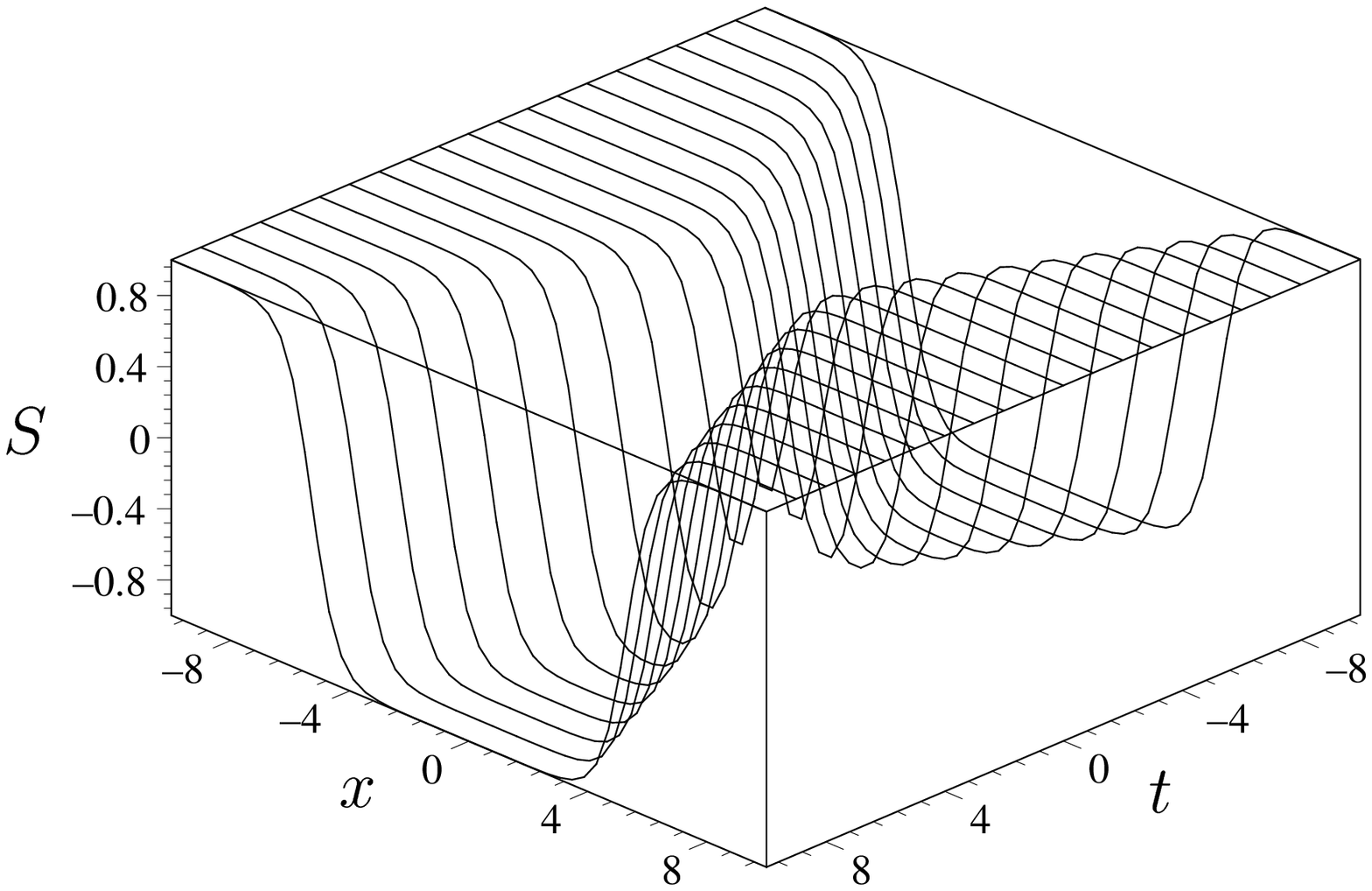,height=8cm,width=6.4cm,angle=270}
\caption{Scalar potential for kink-antikink scattering at $v=0.5$, showing a repulsive interaction.} 
\label{fig1}
\end{center}
\end{figure}

\begin{figure}
\begin{center}
\epsfig{file=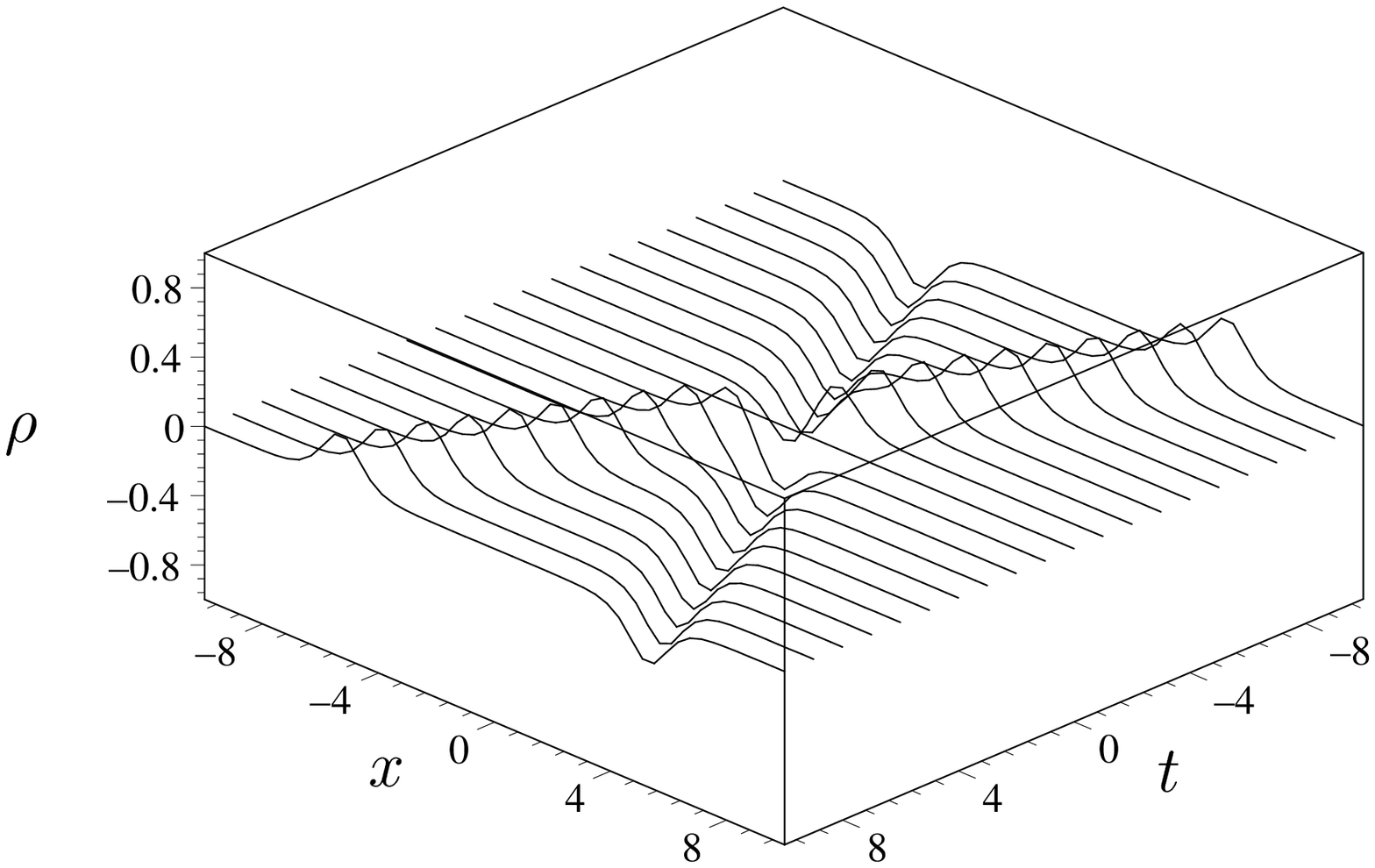,height=8cm,width=6.4cm,angle=270}
\caption{Fermion density for kink-antikink scattering. Parameters: $v=0.5, \nu_1=1, \nu_2=0$ (baryon-antibaryon collision).} 
\label{fig2}
\end{center}
\end{figure}

\begin{figure}
\begin{center}
\epsfig{file=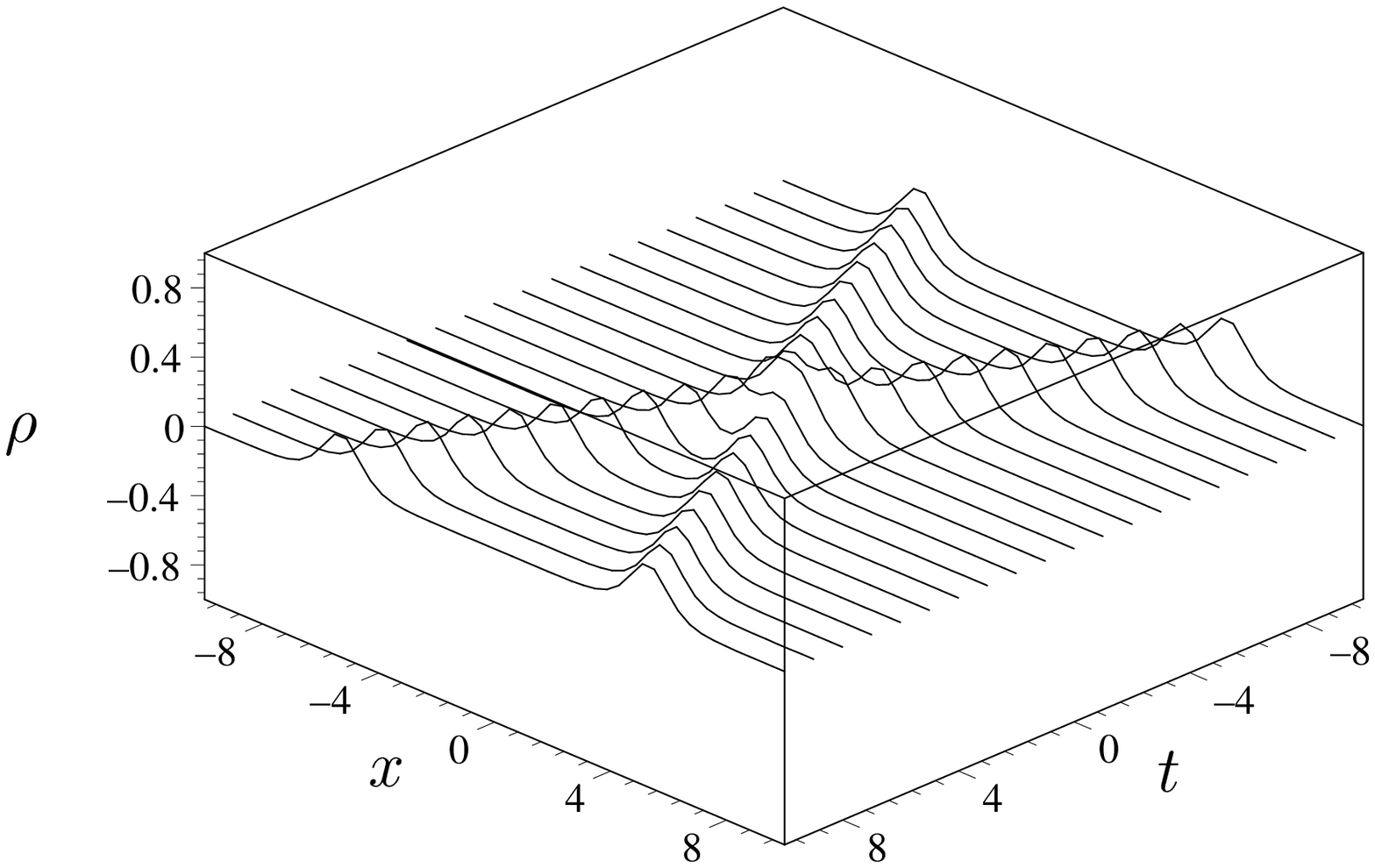,height=8cm,width=6.4cm,angle=270}
\caption{Fermion density for kink-antikink scattering. Parameters: $v=0.5, \nu_1=1, \nu_2=1$ (baryon-baryon collision).} 
\label{fig3}
\end{center}
\end{figure}

We now turn to the kink-antikink scattering problem. Following the suggestion of DHN, we analytically continue the above breather
solution to the value $\epsilon= i/v$. This gives rise to the following changes in our formulae for the potential and the spinors,
\begin{eqnarray}
K & = & \frac{2}{\sqrt{1-v^2}},
\nonumber \\
\Omega & = & -\frac{2 i v}{\sqrt{1-v^2}},
\nonumber \\
\cos \Omega t & = & \cosh \frac{2vt}{\sqrt{1-v^2}},
\nonumber \\
\sin \Omega t & = & - i \sinh \frac{2vt}{\sqrt{1-v^2}},
\nonumber \\
a & = & \sqrt{\frac{1}{v^2}+ \frac{b^2}{1-v^2}}.
\label{13}
\end{eqnarray}
The continuum states now have to be normalized to free spinors at $t\to - \infty$ using
\begin{equation}
{\cal N}_k=\sqrt{\frac{2 k^2}{4 k^2+K^2}}.
\label{15}
\end{equation}
The discrete states are square integrable and normalized to 1 by 
\begin{equation}
{\cal N}_0 =  \sqrt{\frac{K(1+a+b)}{2}}.
\label{16}
\end{equation}
Let us now consider the issue of self-consistency for the scattering problem. The contribution from the negative energy
($\omega<0$) continuum to the scalar potential is found to be 
\begin{eqnarray}
- g^2 \sum_k \bar{\psi}_k \psi_k & = &  S Ng^2  \int_{-\Lambda/2}^{\Lambda/2} \frac{ dk}{2\pi} \frac{1}{\sqrt{k^2+1}} 
\nonumber \\
& &  + f_4 Ng^2  \int_{-\infty}^{\infty}
\frac{ dk}{2\pi} h(k)
\label{17}
\end{eqnarray}
with
\begin{equation}
h(k) = \frac{2bv^2}{\sqrt{1+k^2}(1-v^2)(1+k^2-v^2k^2)}.
\label{18}
\end{equation}
The first term by itself gives the self-consistent result owing to the vacuum gap equation,
\begin{equation}
1 = Ng^2 \int_{-\Lambda/2}^{\Lambda/2} \frac{ dk}{2\pi} \frac{1}{\sqrt{k^2+1}} .
\label{19}
\end{equation}
This observation is the same as for the breather.  Differences arise in the treatment of the discrete states. Since there
are two orthogonal bound states, there is an ambiguity how to fill these states in the TDHF approach. DHN argue that in the case of 
the breather one should take the linear combinations $\psi_0^{(1)}\pm \psi_0^{(2)}$ and fill them with $N$ and $n_0$ fermions, respectively.
We have to reconsider this prescription in the light of the scattering problem. Appropriate initial conditions for the TDHF equation 
are an incoming kink and antikink with prescribed baryon numbers. In order to prepare such an initial state, we need to know
which linear combinations of the two bound states goes over into the kink or antikink (zero-energy) valence state for $t\to - \infty$.
Likewise, in order to be able to interpret the final state of the scattering process, we need the corresponding analysis at
$t\to \infty$.
Consider an arbitrary (normalized) linear combination of $\psi_0^{(1)}, \psi_0^{(2)}$,
\begin{equation}
\psi_0 = \frac{\lambda \psi_0^{(1)}+ \mu \psi_0^{(2)}}{\sqrt{|\lambda|^2+|\mu|^2}}.
\label{20}
\end{equation}
In order to exhibit the asymptotic behavior of this wave function for $x \to \pm \infty, t \to \pm \infty$, we introduce
labels for the incoming/outgoing wave from/to the left/right as follows,
\begin{eqnarray}
{\rm il} & \simeq & t\to - \infty, x \to - \infty
\nonumber \\ 
{\rm ir} & \simeq & t\to - \infty, x \to + \infty
\nonumber \\
{\rm ol} & \simeq & t\to + \infty, x \to - \infty
\nonumber \\
{\rm or} & \simeq & t\to + \infty, x \to + \infty
\label{21}
\end{eqnarray}
Using this shorthand notation, we find the asymptotic expressions (omitting a common normalization factor ${\cal N}_0/4\sqrt{a}\,$)
\begin{eqnarray}
\psi_{0,{\rm il}} & = & \frac{1}{\cosh z_+} \left( \begin{array}{r} \lambda(1-c_4)+ i\mu(c_2+c_3) \\ -  i\lambda(c_2-c_3)+\mu (1+c_4) 
\end{array} \right),
\nonumber \\
\psi_{0,{\rm ir}} & = & \frac{1}{\cosh y_-} \left( \begin{array}{r}   \lambda(1+c_4)-  i\mu(c_2-c_3)      \\     i\lambda (c_2+c_3)+\mu(1-c_4) 
\end{array} \right),
\nonumber \\
\psi_{0,{\rm ol}} & = & \frac{1}{\cosh y_+} \left( \begin{array}{r} \lambda(1-c_4)-  i\mu(c_2+c_3)         \\    i\lambda (c_2-c_3)+\mu(1+c_4)    
   \end{array} \right),
\nonumber \\
\psi_{0,{\rm or}} & = & \frac{1}{\cosh z_-} \left( \begin{array}{r} \lambda(1+c_4)+  i\mu(c_2-c_3)        \\   -  i\lambda (c_2+c_3) + \mu(1-c_4) 
      \end{array} \right),
\label{22}
\end{eqnarray}
with
\begin{eqnarray}
y_{\pm} &=& \frac{1}{2} \left(K(x+vt) \pm \ln a \right) ,
\nonumber \\
z_{\pm} & = & \frac{1}{2} \left(K(x-vt) \pm \ln a \right) .
\label{23}
\end{eqnarray}
We introduce two orthonormal discrete states with parameters ($\lambda_1,\mu_1$) and ($\lambda_2,\mu_2$) and
occupation $N_1,N_2$. In order to match the initial conditions of the kink-antikink scattering problem, we require that state 1 
has only incident fermions from the left and state 2 only incident fermions from the right. Eqs.~(\ref{22}) then imply the two pairs
of homogeneous, linear equations 
\begin{eqnarray}
0 & = & \lambda_1 (1+c_4) -  i\mu (c_2-c_3)
\nonumber \\
0 & = & i\lambda_1(c_2+c_3)+\mu_1(1-c_4)
\nonumber \\
0 & = & \lambda_2(1-c_4)+  i\mu_2 (c_2+c_3)
\nonumber \\
0 & = & -  i\lambda_2(c_2-c_3)+\mu_2(1+c_4)
\label{24}
\end{eqnarray}
The conditions for the existence of a non-trivial solution,
\begin{eqnarray}
0 & = & \det \left( \begin{array}{cc} 1+c_4 & -  i(c_2-c_3) \\ i(c_2+c_3) & 1-c_4 \end{array} \right)  ,
\nonumber \\
0 & = & \det \left( \begin{array}{cc} 1-c_4 & i(c_2+c_3) \\ -  i(c_2-c_3) & 1+c_4 \end{array} \right)  ,
\label{25}
\end{eqnarray}
are indeed satisfied for arbitrary $v,b$. We can then determine the ratios $\mu_i/\lambda_i$,
\begin{eqnarray}
\frac{\mu_1}{\lambda_1} & = & -  i \frac{1+c_4}{c_2-c_3},
\nonumber \\
\frac{\mu_2}{\lambda_2}  & = &  i\frac{1-c_4}{c_2+c_3}.
\label{26}
\end{eqnarray}
The right-hand side is purely imaginary. This implies a vanishing contribution to the chiral condensate from the discrete states
which is proportional to $\lambda^* \mu + \mu^* \lambda$, as can easily be checked. Self-consistency in the scattering problem 
is therefore only possible for $b=0$. In this particular case, the coefficients $c_i$ are 
\begin{equation}
c_2=1, \quad c_3 = - c_4 = - \frac{1-v}{\sqrt{1-v^2}},
\label{26a}
\end{equation}
so that Eqs.~(\ref{26}) simplify to
\begin{equation}
\frac{\mu_1}{\lambda_1} = -  i = - \frac{\mu_2}{\lambda_2}.
\label{27}
\end{equation}
A consistent choice of parameters is therefore $b=0$ and
\begin{equation}
\lambda_1=\lambda_2 = 1, \quad \mu_1 = - \mu_2 = -  i.
\label{28}
\end{equation}
For these parameters, one finds that state 1 has only an incoming wave from the left and an outgoing wave to the right, state 2 only 
an incoming wave from the right and an outgoing wave to the left. The valence fermions are exchanged between 
the two scatterers, presumably a consequence of the fact that the potential is transparent. These valence fermions do not play any role
in the self-consistency issue. We can prescribe the number of valence quarks independently for kink
and antikink as $N_1,N_2$ where $0 \leq N_i \leq N$. In the final state, $N_1,N_2$ are then simply exchanged.
We recall that the kink (or antikink) with $N_i$ valence quarks carries fermion number $N_f=N_i-N/2$ as a result of induced fermion number in
a topologically nontrivial background potential \cite{L19a,L19b}.
Expressed in terms of 
reduced fermion number $N_f/N=N_i/N-1/2=\nu_i-1/2$, the scattering process thus reads
\begin{equation}
K(\nu_2-1/2) + \bar{K}(\nu_1-1/2) \rightarrow K(\nu_1 -1/2) + \bar{K}(\nu_2 - 1/2).
\label{28a}
\end{equation}
The time delay
\begin{equation}
\Delta t = \frac{\ln v}{v}\sqrt{1-v^2}
\label{29}
\end{equation}
is independent of the fermion numbers and negative, indicating a repulsive interaction between kink and antikink.
Note that the value $b=0$ implies that the kink-antikink scattering solution is of type I and hence significantly simpler than the breather. 
This is also reflected in the form of the scalar potential, 
\begin{equation}
S= \frac{v \cosh Kx - \cosh Kvt}{v \cosh Kx+ \cosh Kvt}, \quad K= \frac{2}{\sqrt{1-v^2}}.
\label{30}
\end{equation}
Incidentally, Eq.~(\ref{30}) agrees with a result quoted in Ref.~\cite{L16} without derivation.  
The fermion density for the discrete states is given by
\begin{equation}
\rho_{1,2} = \frac{vK\left[ v+\cosh K(x\pm vt) \right]}{2(v\cosh Kx+ \cosh Kvt)^2},
\label{32}
\end{equation}
where the $+$ sign belongs to state 2 moving to the left, the $-$ sign to state 1 moving to the right, cf. Eq.~(\ref{28}).
Expressions (\ref{32}) have been normalized to 1 and have to be multiplied by the occupation numbers to get the true fermion density.
The density $\psi_k^{\dagger}\psi_k$ of the negative energy continuum states is
\begin{eqnarray}
\rho_k &=& 1 + \frac{2v\left( v+\cosh Kx \cosh Kvt \right)}{(k^2v^2-k^2-1)\left( v \cosh Kx+\cosh Kvt \right)^2}
\label{34} \\
&- &  \frac{2kv^2 \sinh Kx \sinh Kvt}{\sqrt{1+k^2}(k^2v^2-k^2-1)\left(v \cosh Kx+\cosh Kvt\right)^2}.
\nonumber
\end{eqnarray}
Integrating over $k$ and subtracting the vacuum contribution yields the simple, finite result
\begin{equation}
\int_{-\infty}^{\infty}   \frac{ dk}{2\pi} (\rho_k-1) = - \frac{1}{2} (\rho_1+\rho_2).
\label{35} 
\nonumber
\end{equation}
The total fermion density per flavor is thus given by
\begin{equation}
\rho_f  =  \left(\nu_1- \frac{1}{2} \right) \rho_1 +\left(  \nu_2 - \frac{1}{2} \right) \rho_2 .
\label{36} 
\end{equation}
Integration over $x$ finally yields the (conserved) total fermion number,
\begin{equation}
\frac{N_f}{N} = \int_{-\infty}^{\infty} dx \rho_f(x,t) = \nu_1 + \nu_2 -1\ .
\label{37}
\end{equation}
The kink-antikink scattering process is illustrated in Figs.~\ref{fig1}--\ref{fig3}. Fig.~\ref{fig1} shows, by means of the scalar potential $S$, 
that kink and antikink approach each other, are repelled and bounce back. This picture is independent of the baryon numbers involved.
Figs.~\ref{fig2} and \ref{fig3} exhibit the baryon density for baryon-antibaryon and baryon-baryon scattering. In the first case one clearly 
sees the exchange of fermions during the collision.

\section{Effective bosonic field theory}\label{sect3}
A big hurdle in applying the TDHF approach is to find the self-consistent potential. Obviously it would be extremely nice if one could
write down a closed equation satisfied by the potential. This has been achieved approximately in the case of the massive chiral 
GN model near the chiral limit, using the derivative expansion \cite{L13}. Such classical field equations may be regarded as equations of motion
of an effective scalar theory where the fermions have been ``integrated out", in the language of the path integral. The derivative expansion
is of little use in the
discrete chiral GN model where the baryons remain localized in the chiral limit. Here we address the question whether one can 
nevertheless identify an effective bosonic theory for the non-chiral GN model.

Since the tanh kink appears in $\phi^4$ theory, one may be tempted to postulate the equation
\begin{equation}
\partial_{\mu}\partial^{\mu} S + 2 S^3 - 2 S = 0.
\label{39}
\end{equation}
It is indeed solved by the boosted kink $\tanh(\gamma(x-vt))$. However, this cannot be correct because Eq.~(\ref{39}) has no solitons in the
strict sense of the word: If one scatters a kink and an antikink, a numerical study shows that they interact inelastically and do not
keep their $\tanh$ shape \cite{L20}. A simple bosonic theory which does not have this problem can be found for type I solutions only, to which
we
restrict ourselves in this section. These solutions are specific for kink dynamics (kink, kink crystal, kink-antikink scattering), and therefore
only exist in the massless GN model. For type I solutions we can write
\begin{equation}
S = \ell_{\alpha} \bar{\psi}_{\alpha} \psi_{\alpha} 
\label{40}
\end{equation} 
for every occupied state $\alpha$ with nonvanishing $\bar{\psi}_{\alpha}\psi_{\alpha}$. If we pick any such state, the TDHF problem 
evidently reduces to the $N=1$ classical GN model, i.e., a non-linear Dirac equation 
\begin{equation}
\left( i \partial \!\!\!/ - \ell_{\alpha} \bar{\psi}_{\alpha} \psi_{\alpha} \right) \psi_{\alpha} = 0
\label{41}
\end{equation}
with $c$-number spinors. We suppress the label $\alpha$ from now on. Neveu and Papanicolaou \cite{L16} have proven
the integrability of the classical $N=1,2$ GN models long ago. The part of their work dealing with $N=1$ in fact contains the answer
to the question about the effective bosonic theory. To demonstrate this fact we closely follow their work in the present section.

To this end it is useful to switch to a different representation of the $\gamma$-matrices in which the upper and lower components of the
Dirac spinor $\psi_1, \psi_2$ have definite chirality,
\begin{equation}
\gamma^0 = \sigma_1, \quad \gamma^1 = i \sigma_2, \quad \gamma_5 = - \sigma_3 \ .
\label{42}
\end{equation}
Using light-cone coordinates
\begin{equation}
z=x-t, \quad \bar{z} = x+t
\label{43}
\end{equation}
and the comma-notation for partial derivatives, the non-linear Dirac equation assumes the simple form
\begin{eqnarray}
-2i\psi_{1,z} & = & S \psi_2,
\nonumber \\
2i \psi_{2,\bar{z}} & = & S \psi_1,
\label{44}
\end{eqnarray}
where
\begin{equation}
S = \ell \left( \psi_1^*\psi_2+ \psi_2^* \psi_1 \right).
\label{45}
\end{equation}
The Dirac equation expresses $\psi_{1,z}$ and $\psi_{2,\bar{z}}$ in terms of $\psi_1,\psi_2$. What about the other derivatives,
$\psi_{1,\bar{z}}$ and $\psi_{2,z}$? Following Neveu and Papanicolaou, one first derives the identities
\begin{eqnarray}
S \psi_{1,\bar{z}}- S_{,\bar{z}}\psi_1 & = & -i h_1 \ell \psi_2,
\nonumber \\
S \psi_{2,z}-S_{,z}\psi_2 & = & -i h_2 \ell \psi_1,
\label{46}
\end{eqnarray}
with 
\begin{eqnarray}
h_1 & = & i \left( \psi_1^* \psi_{1,\bar{z}} - \psi_1 \psi_{1,\bar{z}}^* \right),
\nonumber \\
h_2 & = &  i \left( \psi_2^* \psi_{2,z} - \psi_2 \psi_{2,z}^* \right).
\label{47}
\end{eqnarray}
Since the right-hand side depends again on the unknown derivatives $\psi_{1,\bar{z}},\psi_{2,z}$ and their complex conjugates, it looks
as if the goal of expressing these derivatives through $\psi_1,\psi_2$ had not been reached. However, simple algebra shows that
\begin{equation}
h_{1,z} = 0 = h_{2,\bar{z}}
\label{48}
\end{equation}
so that $h_{1,2}$ can only depend on either $z$ or $\bar{z}$. If these functions would not be constant, they would describe fermion bilinears
propagating at the speed of light. In the absence of massless particles in the model we conclude that $h_1,h_2$ are constant
for physically sensible solutions. This enables us to express all derivatives of $\psi$ in terms of $\psi$ itself as follows,
\begin{equation}
\psi_{,\bar{z}} = C_1 \psi, \qquad \psi_{,z} = C_2 \psi,
\label{49}
\end{equation}
with the matrices
\begin{eqnarray}
C_1 & = &  \left( \begin{array}{cc} S_{,\bar{z}}S^{-1} & -i h_1 \ell S^{-1} \\ -i S/2 & 0 \end{array} \right), 
\nonumber \\
C_2 & = &  \left( \begin{array}{cc} 0  & i S/2 \\ -ih_2 \ell S^{-1} & S_{,z} S^{-1} \end{array} \right).
\label{50}
\end{eqnarray}
The integrability condition for the system (\ref{49}) reads
\begin{equation}
C_{1,z} - C_{2,\bar{z}} + \left[ C_1,C_2 \right] = 0
\label{51}
\end{equation}
or equivalently
\begin{equation}
S S_{,z \bar{z}} - S_{,z}S_{,\bar{z}} - \frac{1}{4} S^4 = h_1 h_2 \ell^2
\label{52}
\end{equation}
As the right-hand side is constant, we have succeeded in deriving a non-linear partial differential equation for
the self-consistent TDHF potential, at least for type I solutions. If we disregard the kink crystal for the moment and focus
on localized solutions where $S$ reaches its vacuum value asymptotically, we can even determine the constant.
In units where $S=m=1$ in the vacuum, we infer from Eq.~(\ref{52}) that
\begin{equation}
h_1 h_2 \ell^2 = - \frac{1}{4}\ .
\label{53}
\end{equation}
In normal coordinates, the final equation for $S$ reads
\begin{equation}
S \partial_{\mu} \partial^{\mu}S - \partial_{\mu}S \partial^{\mu}S + S^4-1 = 0.
\label{53a}
\end{equation}
It is satisfied by the kink and the kink-antikink solutions in arbitrary Lorentz frames, but also by the kink crystal for which we have not derived
the value of the constant, Eq.~(\ref{53}). One can write down a simple (although singular) Lagrangian which yields Eq.~(\ref{53a}) as
Euler-Lagrange equation, namely
\begin{equation}
{\cal L} = \frac{1}{S^2} \left( \partial_{\mu}S \partial^{\mu}S - S^4 -1 \right).
\label{53b}
\end{equation}
However, this should not be interpreted as effective bosonic field theory for the GN model. If one derives the Hamiltonian density 
${\cal H}$ from ${\cal L}$ in the usual way, one does not get the correct energy density. The reason is presumably the fact that we already
made use of properties of the solution when deriving Eq.~(\ref{53a}), notably the fact that we are dealing with a type I solution of the TDHF 
problem. In this way we are not really able to integrate out the fermions in full generality and construct the effective bosonic action
for arbitrary scalar fields.

Eq.~(\ref{53a}) does not yet resemble any of the well-known equations with solitonic solutions. 
The closest we could come to a more familiar looking form was by means of the change of variables \cite{L16}
\begin{equation}
S^2=e^{\theta}, \qquad \theta= \ln S^2,
\label{54}
\end{equation}
which reduces Eq.~(\ref{53a}) to the sinh-Gordon equation,
\begin{equation}
\partial_{\mu} \partial^\mu \theta + 4  \sinh \theta = 0.
\label{55}
\end{equation}
However we loose the information about the sign of $S$ and hence also the Z$_2$ chiral symmetry in this nonlinear transformation -- 
the two vacua $S=\pm 1$ are mapped onto the same value $\theta=0$. Since the zero's of $S$ give rise to singularities of $\theta$, it is easy 
to reconstruct a solution $S$ of Eq.~(\ref{53a}) from a singular solitonic solution $\theta$ of the sinh-Gordon equation, so that the mapping 
is nevertheless quite useful.  With this caveat, the matrices $C_1,C_2$ and the linear equations (\ref{49}) can then be identified with the 
Lax pair of the sinh-Gordon equation. 

Notice that the coefficient 4 in Eq.~(\ref{55}) has a simple physical interpretation: The linearized sinh-Gordon equation, 
\begin{equation}
\left( \partial_{\mu} \partial^{\mu}  + 4 \right) \theta = 0,
\label{56}
\end{equation}
yields the Klein-Gordon equation for a scalar meson with mass 2 in units of the fermion mass. This can be identified with the $\sigma$ meson 
of the massless GN model. The relation between kink, $\sigma$ meson and sinh-Gordon equation in the massless
GN model is analogous to the relation between the light baryon, the $\pi$ meson and the sine-Gordon equation in
the massive 2d Nambu--Jona-Lasinio model (NJL$_2$) close to chiral limit (derivative expansion \cite{L13}). In the latter case, this was
interpreted as the simplest
realization of the Skyrme model in the sense that baryons are topologically non-trivial excitations of the pion field \cite{L21}. In the GN
model, we have now identified a similar picture for the case of a discrete chiral symmetry, the baryon emerging as a large amplitude 
excitation of the $\sigma$ field.
The similarity 
between the baryons in both cases is particularly striking if one writes the sine-Gordon and sinh-Gordon solitons in the following form:
\begin{enumerate}
\item
Massless GN model (exact, $m_{\sigma}=2$),
\begin{eqnarray}
S^2 & = &  e^{\theta}, \qquad \theta \ =\  - 4 {\rm \ artanh\ }e^{-m_{\sigma}x},
\nonumber \\
0 & = & \partial_{\mu}\partial^{\mu} \theta + m_{\sigma}^2 \sinh \theta  .
\label{59}
\end{eqnarray}
\item
Massive NJL$_2$ model (leading order derivative expansion, $m_{\pi}=2\sqrt{\gamma}$ with $\gamma$ the confinement parameter \cite{L13}),
\begin{eqnarray}
S-iP & = & e^{i\phi}, \qquad \phi \ =\  4 \arctan e^{m_{\pi}x},
\nonumber \\
0 & = &  \partial_{\mu}\partial^{\mu} \phi + m_{\pi}^2 \sin \phi.  
\label{60}
\end{eqnarray}
\end{enumerate}
We can also write down a common formula for topological baryon number,
\begin{equation}
\ln \Phi(\infty) - \ln \Phi(-\infty) = 2\pi i N_B,
\label{61}
\end{equation}
with $\Phi=S$ for the GN model and $\Phi=S-iP$ for the NJL$_2$ model. However, in the GN case, this only determines the non-integer part
of the induced baryon number \cite{L22} (kink and antikink give $\mp 1/2$ although they have the same value of induced baryon number of 
$-1/2$) so that the analogy should perhaps not be overrated. 
 
The kink belongs to the class of ``traveling wave solutions" of the sinh-Gordon equation. Kink-antikink scattering
is an example of a  ``functional separable solution" \cite{L23}
\begin{equation}
\theta(x,t)  =   4 \,{\rm artanh} \left[ f(t)g(x) \right]
\label{57}
\end{equation}
where
\begin{eqnarray}
(f_{,t})^2  & = &   A f^4 + B f^2 + C,
\nonumber \\
- (g_{,x})^2 & = &  C g^4 + (B+4)g^2 + A.
\label{58}
\end{eqnarray}
The general $N$ soliton solution is also known for the sinh-Gordon equation  \cite{L19,L24} and is a likely candidate for the 
TDHF solution of the GN model with $N$ kinks and antikinks. Since the Lax pair including the spinor wave functions are
known, all what one would have to do is find the bound state solutions and verify self-consistency. From the point of view of
particle physics, this is perhaps somewhat academic because it describes a scattering problem with $N$ incident baryons. Nevertheless, it 
would be challenging to solve the relativistic $N$-baryon problem exactly and analytically in terms of the elementary fermion constituents,
including the Dirac sea effects.
\section{Relation to strings in AdS$_3$}\label{sect4}
Jevicki et al. have explored the close relationship between sinh-Gordon theory on the one hand and classical strings in AdS$_3$ on the other 
hand \cite{L17,L18,L19}. Since the GN model can also be related to the sinh-Gordon model, this should enable us to map type I solutions of the 
GN model onto solutions of
classical string theory. From the physics point of view, a stringy interpretation is not obvious for a non-gauge theory like the GN model. 
It may be of interest to see explicitly how such a mapping works.

Let us briefly review the required string theory background  following Ref.~\cite{L17}. The AdS$_3$ target space is parametrized by the
embedding coordinates $Y_a (a=-1,0,1,2)$ in R$^{2,2}$,
\begin{equation}
\vec{Y} \cdot \vec{Y} := - Y_{-1}^2-Y_0^2+Y_1^2+Y_2^2 = -1.
\label{62}
\end{equation}
The string equation of motion in conformal gauge reads ($\partial=\partial_z, \bar{\partial}=\partial_{\bar{z}}$)
\begin{equation}
\partial {\bar{\partial}} \vec{Y} - ( \partial \vec{Y} \cdot \bar{\partial} \vec{Y}) \vec{Y}=0
\label{63}
\end{equation}
and has to be supplemented by the Virasoro constraints
\begin{equation}
( \partial \vec{Y} )^2 = 0 =( \bar{\partial}\vec{Y} )^2 .
\label{64}
\end{equation}
The crucial instrument in relating strings in AdS$_3$ to the sinh-Gordon equation is the Pohlmeyer reduction \cite{L25}, using the
factorization of the AdS$_3$ isometry group SO(2,2) into SO(2,1)$\times$SO(2,1).
The string coordinates are expressed through two auxiliary spinors $\phi,\chi$
as follows,
\begin{eqnarray}
Z_1 & = & Y_{-1} + i Y_0 \ = \ \phi_1^*\chi_1 - \phi_2^*\chi_2,
\nonumber \\
Z_2 & = & Y_1+iY_2 \ = \ \phi_2^*\chi_1^* - \phi_1^*\chi_2^*.
\label{65}
\end{eqnarray}
Here, $\phi$ and $\chi$ are normalized according to
\begin{equation}
1 = \phi_1^*\phi_1-\phi_2^*\phi_2 = \chi_1^*\chi_1-\chi_2^*\chi_2
\label{66}
\end{equation}
and satisfy the linear system of equations
\begin{eqnarray}
\phi_{,\bar{z}} & = & A_1 \phi, \qquad \phi_{,z} \ = \  A_2 \phi,
\nonumber \\
\chi_{,\bar{z}} & = & B_1 \chi, \qquad \chi_{,z} \ = \ B_2 \chi,
\label{67}
\end{eqnarray}
where now $z,\bar{z}$ are light-cone coordinates derived from worldsheet space-time coordinates. 
The matrices depend on two arbitrary functions $u(\bar{z}),v(z)$ which we choose as $u=2,v=-2$ for the sake of simplicity.
For these particular values the normalization of the light-cone coordinates agrees with Eq.~(\ref{43}) if we identify $x,t$ with the worldsheet
coordinates. The matrices introduced in Eq.~(\ref{67}) are then given by
\begin{eqnarray}
A_1 & = & \frac{1}{4} \left( \begin{array}{cc} -i\lambda c_+ & i \alpha_{,\bar{z}} - \lambda c_- \\ -i \alpha_{, \bar{z}} - \lambda c_- & i \lambda c_+ 
    \end{array}\right) ,
\nonumber \\ 
A_2 & = &  \frac{1}{4}\left( \begin{array}{cc}  i c_+/\lambda & -i \alpha_{,z}- c_-/\lambda \\ i \alpha_{,z}-  c_-/\lambda  &  -i c_+ /\lambda   \end{array}
\right),
\nonumber \\
B_1 & = & \frac{1}{4} \left( \begin{array}{cc} -i\lambda c_- & i \alpha_{,\bar{z}} - \lambda c_+ \\ -i \alpha_{, \bar{z}} - \lambda c_+ & i \lambda c_-  
   \end{array}\right) ,
\nonumber \\
B_2  & = & \frac{1}{4}\left( \begin{array}{cc}  i c_-/\lambda & -i \alpha_{,z}- c_+/\lambda \\ i \alpha_{,z}-  c_+/\lambda  &  -i c_- /\lambda   \end{array}
\right),
\label{68}
\end{eqnarray}
with
\begin{equation}
c_{\pm} = e^{-\alpha/2} \pm e^{\alpha/2}.
\label{69}
\end{equation}
$\alpha(z,\bar{z})$ is a function which is expressible in terms of $\phi$, but this relation will not be needed here. 
The integrability condition for both systems of linear equations (\ref{67}) yields the sinh-Gordon equation,
\begin{equation}
\alpha_{,z \bar{z}} = \sinh \alpha.
\label{70}
\end{equation}
It is straightforward to verify that a pair of spinors $\phi,\chi$ satisfying the normalization conditions (\ref{66}) and the linear sytem
 (\ref{67}) yields
a solution of the string equations (\ref{62}--\ref{64}), the link between the two problems being given by Eqs.~(\ref{65}).

The way to relate the GN model to strings in AdS$_3$ is to find a gauge transformation between two different representations
of the sinh-Gordon Lax pair, Eqs.~(\ref{49},\ref{50}) and Eqs.~(\ref{67},\ref{68}), where we have to expect complications due to the
fact that the mapping from the GN model to the sinh Gordon model was not one-to-one. Using the parametrization
\begin{equation}
S = \pm e^{\theta/2}
\label{70a}
\end{equation}
depending on the sign of $S$, let us first consider the case $S>0$.
The gauge transformation from
($C_1,C_2,\psi$) to ($A_1,A_2,\phi$) is given by 
\begin{eqnarray}
A_1 & = & \Omega \left( C_1 - \bar{\partial} \right) \Omega^{-1},
\nonumber \\
A_2 & = & \Omega \left( C_2 - \partial \right) \Omega^{-1},
\nonumber \\
\phi & = & \Omega \psi.
\label{71}
\end{eqnarray}
$A_{1,2}$ depend on $\alpha$ and a scale parameter $\lambda$, $C_{1,2}$ depend on $S=e^{\theta/2}$ and the constants $h_1\ell, h_2\ell$ 
related
through Eq.~(\ref{53}). For future convenience, we introduce a parameter $\zeta$ (to be identified with the spectral parameter of the GN model
TDHF solutions) via 
\begin{equation}
h_1 \ell = 2 \zeta^2, \quad h_2 \ell = - \frac{1}{8 \zeta^2}.
\label{72}
\end{equation}
$\Omega$ can be found in 2 successive steps: First, make a local  Abelian gauge transformation to render $C_1,C_2$ traceless ($A_1,A_2$
are in the SO(2,1) Lie algebra). This is achieved by the choice
\begin{equation}
\Omega_1 = S^{-1/2}.
\label{73}
\end{equation}
Secondly, perform a global non-Abelian gauge transformation
\begin{equation}
\Omega_2 = \left( \begin{array}{cc} \lambda^{-1} & 1 \\ i\lambda^{-1}  & -i \end{array} \right) .
\label{74}
\end{equation}
The product 
\begin{equation}
\Omega= \Omega_2 \Omega_1 =  \left( \begin{array}{cc} \lambda^{-1} & 1 \\ i\lambda^{-1}  & -i \end{array} \right) S^{-1/2} \qquad (S>0)
\label{75}
\end{equation}
then transforms $C_1,C_2$ onto $A_1,A_2$ provided we relate the field variables ($\theta, \alpha$)
and spectral parameters ($\zeta, \lambda$) as follows,
\begin{equation}
\alpha = - \theta, \qquad \lambda^2 = 4 \zeta^2.
\label{76}
\end{equation}
As will be seen later on in concrete examples, the correct sign for negative energy continuum states is $\lambda=-2 \zeta$. Hence the
mapping from the GN solution $\psi$ to the first string $\sigma$ model spinor $\phi$ is
\begin{equation}
\phi = \Omega  \psi(\zeta=- \lambda/2).
\label{77}
\end{equation}
The normalization condition (\ref{66}) then tells us what the correct normalization of the GN model spinor (which differs from the one used in
the standard TDHF approach) should be, namely
\begin{equation}
1 = \phi^{\dagger} \sigma_3 \phi = \frac{2}{\lambda S} \bar{\psi}\psi =  - \frac{1}{\zeta S} \bar{\psi}\psi.
\label{78}
\end{equation}
Comparison with Eq.~(\ref{45}) shows that $\ell=- 1 / \zeta$.
We now turn to the second spinor $\chi$ and the gauge transformation from $C_{1,2}$ to $B_{1,2}$. The same gauge transformation
$\Omega$ can 
be used once again, but Eq.~(\ref{76}) is replaced by
\begin{equation}
\alpha = - \theta, \qquad \lambda^2 = - 4 \zeta^2.
\label{79a}
\end{equation}
Since $\lambda$ on the string side is real, we have to analytically continue the TDHF spinors to purely imaginary spectral parameter 
$\zeta= \pm i \lambda/2$. Which sign is the correct one? It turns out 
that the normalization condition (\ref{66}) for $\chi$ can only be satisfied by taking a linear combination of
spinors with both signs \cite{L19}. The norm vanishes if we keep only one sign. We choose
\begin{equation}
\chi =\Omega \frac{1}{\sqrt{2}} \left[ \psi(\zeta=- i\lambda/2) +
i \psi(\zeta=i \lambda/2) \right]
\label{80}
\end{equation}
where $\psi(\zeta)$ is normalized according to (\ref{78}) before the analytic continuation. The detailed justification is given in the appendix.
These algebraic manipulations are rooted in the symmetries of the GN model on the one hand and AdS$_3$ space on the other hand. 
Neveu and Papanicolaou have identified a dynamical SO(2,1) symmetry of the classical $N=1$ GN model which is also relevant for
type I solutions of the large $N$ quantum theory. To match the SO(2,1)$\times$SO(2,1) symmetry of AdS$_3$, one needs two independent 
spinor solutions -- this is the non-trivial part of the mapping. One of them is given directly by the (appropriately normalized)
TDHF solution, the other involves an analytic continuation of the first one to imaginary spectral parameters. This is apparently what it
takes to embed the fermionic quantum field theory in a higher dimensional, classical string theory.
We can now write down the string coordinates in compact form. Introduce the basic GN TDHF spinor and the 
analytically continued one as [see Eqs.~(\ref{77},\ref{80})]
\begin{eqnarray}
\psi_a & = & \Omega^{-1} \phi \ = \ \psi (\zeta=-\lambda/2),
\label{81} \\
\psi_b & = &  \Omega^{-1} \chi = \frac{1}{\sqrt{2}} \left[ \psi(\zeta=-i\lambda/2) + i \psi(\zeta=i \lambda/2) \right].
\nonumber
\end{eqnarray}
normalized according to (cf. Eq.~(\ref{78}) and the appendix)
\begin{equation}
\bar{\psi}_a \psi_a = \bar{\psi}_b \psi_b =  \frac{\lambda S}{2}.
\label{82}
\end{equation}
Then, using the above gauge transformation,  the string coordinates (\ref{65}) can be expressed in the concise form
\begin{equation}
Z_1 =  \frac{\bar{\psi}_a\psi_b}{\bar{\psi}_a \psi_a}, \quad Z_2 = -  \frac{\bar{\psi}_a i \gamma_5 \psi_b^*}{\bar{\psi}_a \psi_a}.
\label{83}
\end{equation}
Here it looks as if the normalization of the TDHF spinors would simply drop out, provided we use $\bar{\psi}_a\psi_a = \bar{\psi}_b\psi_b$.
However Eq.~(\ref{82}) for $\psi_a$ is important 
because it teaches us how to normalize the GN spinor before doing the analytic continuation. Otherwise, the whole procedure would be ill
defined, since the normalization factor will in general depend on $\zeta$.

All of these manipulations were done under the tacit assumption $S>0$. If $S<0$, we can repeat the same procedure with the following 
changes. 
The gauge transformation defined in Eq.~(\ref{75}) has to be replaced by
\begin{equation}
\Omega=  \left( \begin{array}{cc} \lambda^{-1} & -1 \\ i\lambda^{-1}  & i \end{array} \right) (-S)^{-1/2} \qquad (S<0).
\label{83a}
\end{equation} 
One can easily check that all other equations, in particular the final result (\ref{83}), then remain valid.

Using Eqs.~(\ref{82},\ref{83}) it is straightforward to verify that the string equations (\ref{62}--\ref{64}) are satisfied. One needs the
fact that $\psi_a,\psi_b$ satisfy the Lax system (\ref{49},\ref{50}) with the parameters $h_1\ell=\lambda^2/2$ for $\psi_a$ and
$h_1\ell=- \lambda^2/2$ for $\psi_b$; $h_2 \ell$ then follows from (\ref{53}). The AdS$_3$ condition (\ref{62}) can be shown via the Fierz 
identity
\begin{equation}
 - |\bar{\psi}_a \psi_b |^2 + |\bar{\psi}_a  \gamma_5 \psi_b^*|^2 = - \bar{\psi}_a\psi_a \bar{\psi}_b\psi_b.
\label{85}
\end{equation}  
The Virasoro constraints (\ref{64}) follow trivially from the useful identities
\begin{eqnarray}
Z_{1,z} & = & - \frac{2i}{\lambda^3S^2} \psi_{a1}^*\psi_{b1}, \quad Z_{2,z} \ = \ \frac{2}{\lambda^3 S^2} \psi_{a1}^*\psi_{b1}^*
\nonumber \\
Z_{1,\bar{z}} & = & \frac{2i \lambda}{S^2} \psi_{a2}^* \psi_{b2}, \quad Z_{2,\bar{z}} \ = \ \frac{2\lambda}{S^2} \psi_{a2}^*\psi_{b2}^*
\label{86}
\end{eqnarray}
Finally, the equation of motion (\ref{63}) can be shown with the help of
\begin{equation}
\partial \vec{Y} \cdot \bar{\partial}\vec{Y} = \frac{1}{2S^2}.
\label{87}
\end{equation}
Let us illustrate this mapping between type I solutions of the GN model and classical strings in AdS$_3$ by means of examples
involving zero, one and two solitons, respectively.
\begin{figure}
\begin{center}
\epsfig{file=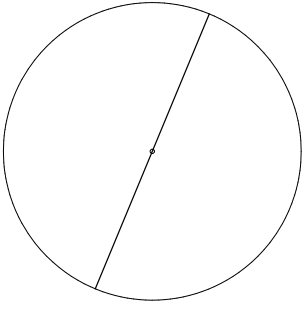,height=4cm,width=4cm,angle=270}
\caption{String from zero soliton TDHF solution (vacuum) in AdS$_3$ for fixed global time. The coordinates on the Poincare disk are given in
Eq.~(\ref{91}). The string is spinning rigidly counterclockwise around its center. }
\label{fig4}
\end{center}
\end{figure}

{\em Zero soliton solution.} In the vacuum, $S=m=1$ and the correctly normalized HF spinor in the present representation reads
\begin{equation}
\psi(\zeta) = \left( \begin{array}{c} \zeta \\ -1/2 \end{array} \right) e^{i(\bar{z} \zeta - z/4\zeta)}, \qquad \zeta= (k-\omega)/2.
\label{88}
\end{equation}
The spectral parameter $\zeta$ is closely related to the light-cone momentum or energy, depending on conventions. 
The string coordinates (\ref{83}) are given by 
\begin{eqnarray}
Z_1 & = &  \sqrt{i} e^{iA_-} \cosh A_+,
\nonumber \\
Z_2 & = &  \sqrt{-i} e^{iA_-} \sinh A_+,
\nonumber \\
\quad A_{\pm} & = &  \frac{\bar{z}\lambda}{2} \pm \frac{z}{2\lambda}.
\label{89}
\end{eqnarray}
To exhibit the motion of the string, we first introduce global coordinates in AdS$_3$ through
\begin{equation}
Z_1 = e^{it} \cosh \xi, \qquad Z_2 = e^{i\phi} \sinh \xi
\label{90}
\end{equation}
where $t$ is the global time and ($\xi,\phi$) are coordinates on the time slice, the hyperbolic plane H$_2$. 
The string at fixed $t$ is most easily visualized on the Poincare disk of radius 1 by the choice of coordinates 
\begin{equation}
X= \rho \cos \phi, \quad Y = \rho \sin \phi, \quad \rho = \sqrt{\frac{\cosh \xi -1}{\cosh \xi +1}}. 
\label{91}
\end{equation}
The circumference of the disk is the boundary of AdS$_3$ space.
Since we have two independent variables $z, \bar{z}$ while keeping only $t$ fixed, Eqs.~(\ref{91}) define a one-dimensional curve. In the case of the vacuum,
the string is evidently just a straight line along a diameter of the disk uniformly spinning around its center, see Fig.~\ref{fig4}.
This agrees with the findings of \cite{L17} where the sinh-Gordon vacuum was mapped onto string solutions.   
\begin{figure}
\begin{center}
\epsfig{file=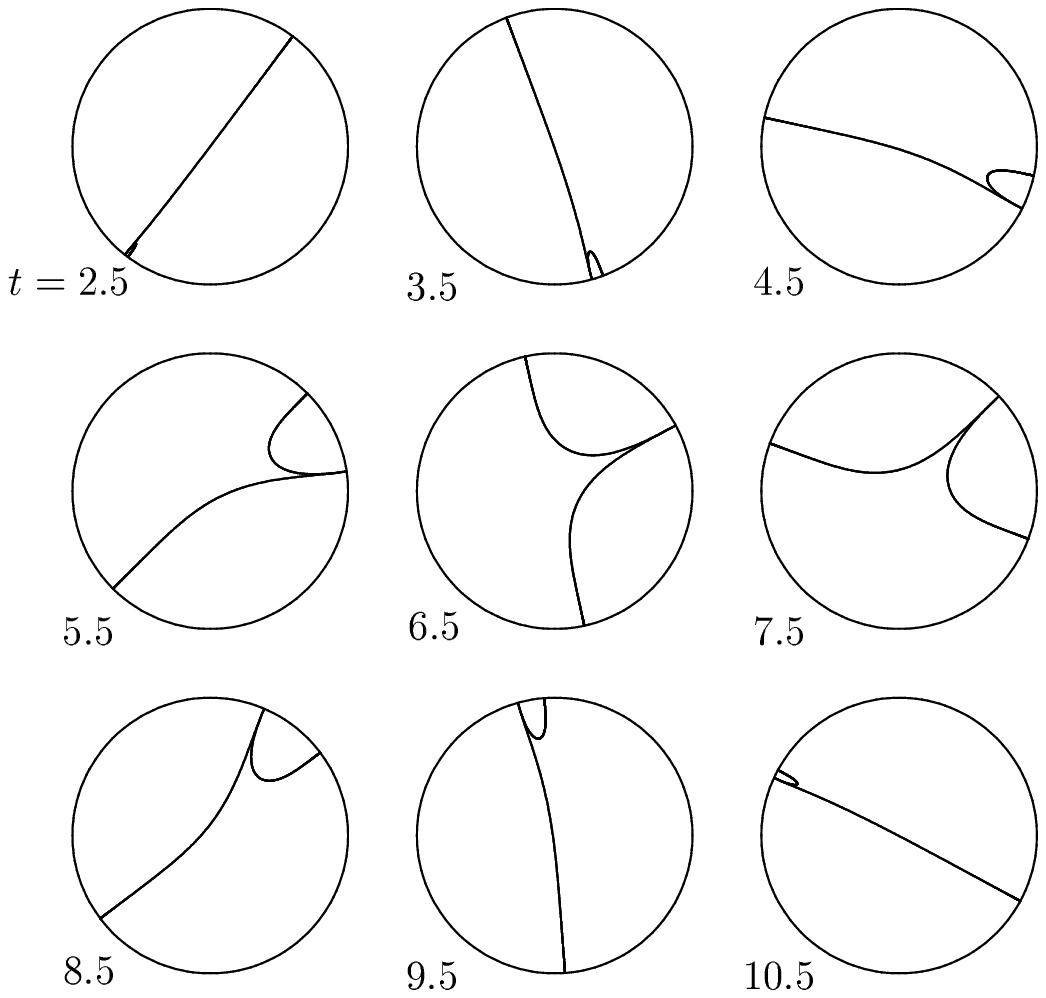,height=8cm,width=7.5cm,angle=270}
\caption{Motion of the string for the one-soliton TDHF solution (kink at rest, $\lambda=-2$) in AdS$_3$. The parameter $t$ is the global time
according to the parametrization (\ref{90}). The spike corresponds to the zero of $S$ and must lie on the boundary of the disk.} 
\label{fig5}
\end{center}
\end{figure}

{\em One soliton solution.} A kink moving with velocity $v$ has the scalar potential 
\begin{equation}
S(x,t) = \tanh \gamma (x-vt), \qquad \gamma= \frac{1}{\sqrt{1-v^2}}.
\label{92}
\end{equation}
The upper and lower components of the continuum spinors belonging to this potential are
\begin{eqnarray}
\psi_1 &=& {\cal N} \left( 1+ \frac{i Y S}{2\zeta} \right) e^{i{\cal A}},
\nonumber \\
\psi_2 & = & {\cal N} \left( -  \frac{ iY}{4 \zeta^2}- \frac{S}{2\zeta}\right)  e^{ i{\cal A}},
\label{93}
\end{eqnarray}
with
\begin{eqnarray}
{\cal A} & = & kx-\omega t \ = \ \zeta \bar{z}-\frac{z}{4\zeta},
\nonumber \\
{\cal N} & = & - \frac{2 i\zeta^2}{Y-2 i\zeta},
\nonumber \\
Y & = &  \sqrt{\frac{1-v}{1+v}}.
\label{94}
\end{eqnarray}
Here, the normalization 
\begin{equation}
\bar{\psi} \psi = - \zeta S
\label{95}
\end{equation}
has been chosen in accordance with Eq.~(\ref{78}).
$Z_1$ and $Z_2$ can be constructed using Eqs.~(\ref{81}--\ref{83}) above, but the result is not very transparent and will not be given here.
In order to map out the motion of the string, we proceed as follows: After choosing an initial global time $t$, we construct a trajectory in the ($z,\bar{z}$) plane
by numerically solving the transcendental equation
\begin{equation}
t = \arctan \frac{Y_{0}}{Y_{-1}}.
\label{96}
\end{equation}
Along this trajectory we then evaluate $Z_1,Z_2$ and convert them to the coordinates $(X,Y)$ according to Eqs.~(\ref{90},\ref{91}). We then plot the string
and repeat the procedure for a sequence of time steps. We recall that we had to choose different gauge transformations depending on the sign of $S$, 
cf. Eqs.~(\ref{75}), (\ref{83a}). When gluing together the corresponding solutions, we impose continuity on the string coordinates in order to
specify the relative phase between the spinors $\psi_a,\psi_b$, which would otherwise be undetermined. In this way we find that we have to change the sign of 
$Z_1,Z_2$ whenever $S$ crosses zero. Fig.~\ref{fig5} shows the result of such a calculation, using the parameters $v=0$ (kink at rest)
and $\lambda=-2$. As expected, the endpoints of the string lie on the boundary.
Since $S=0$ is a singular point, the fold in the string also touches the boundary, unlike in Ref.~\cite{L17}.
This behavior can easily be traced back to the factor $1/\sqrt{|S|}$ in the gauge transformation $\Omega$, see Eqs.~(\ref{75}) and (\ref{83a}).
Comparing Figs.~\ref{fig4} and \ref{fig5} we observe that the kink string resembles the vacuum string at the beginning and the end of the motion shown, reflecting
the characteristic behavior of the kink potential which also interpolates between two vacua. In order to better understand the shape of the strings, we
also recall that geodesics on the Poincare disk are either straight lines through the center or circles intersecting the boundary at right angles. 

{\em Two soliton solution.} It is straightforward to repeat such a calculation for the kink-antikink scattering solution discussed above, at least using Maple.
Since $S$ now changes sign twice, the string is obtained by patching together three different solutions. The resulting string now connects two points 
on the boundary of the disk and touches this boundary in cusps at two intermediate points corresponding to the zeros of $S$. Figs.~\ref{fig6},\ref{fig7} show an
example derived from kink-antikink scattering at $v=\tanh 1 \approx 0.76$ (rapidity 1) and $\lambda=-2$. Since the motion is already fairly complicated, we have 
chosen finer time steps than for the one soliton case. Each of the three parts of the string is infinitely long due to the Poincare metric and qualitatively follows
the geodesics.

\begin{figure}
\begin{center}
\epsfig{file=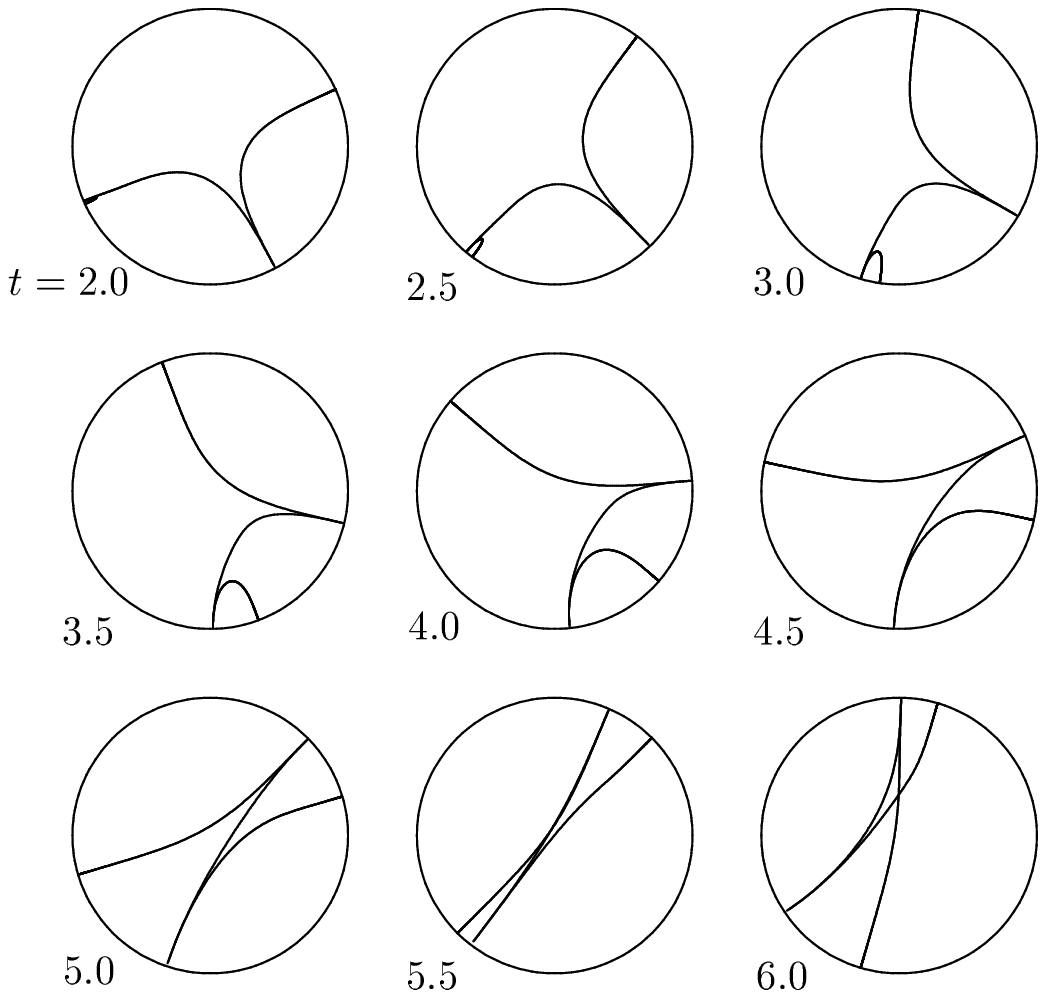,height=8cm,width=7.5cm,angle=270}
\caption{Time evolution of string derived from kink-antikink scattering at $v=0.76, \lambda=-2$. The representation is the same as in Fig.~\ref{fig5}, but 
now there are two cusps on the boundary, reflecting the two zeros of $S$.} 
\label{fig6}
\end{center}
\end{figure}
\begin{figure}
\begin{center}
\epsfig{file=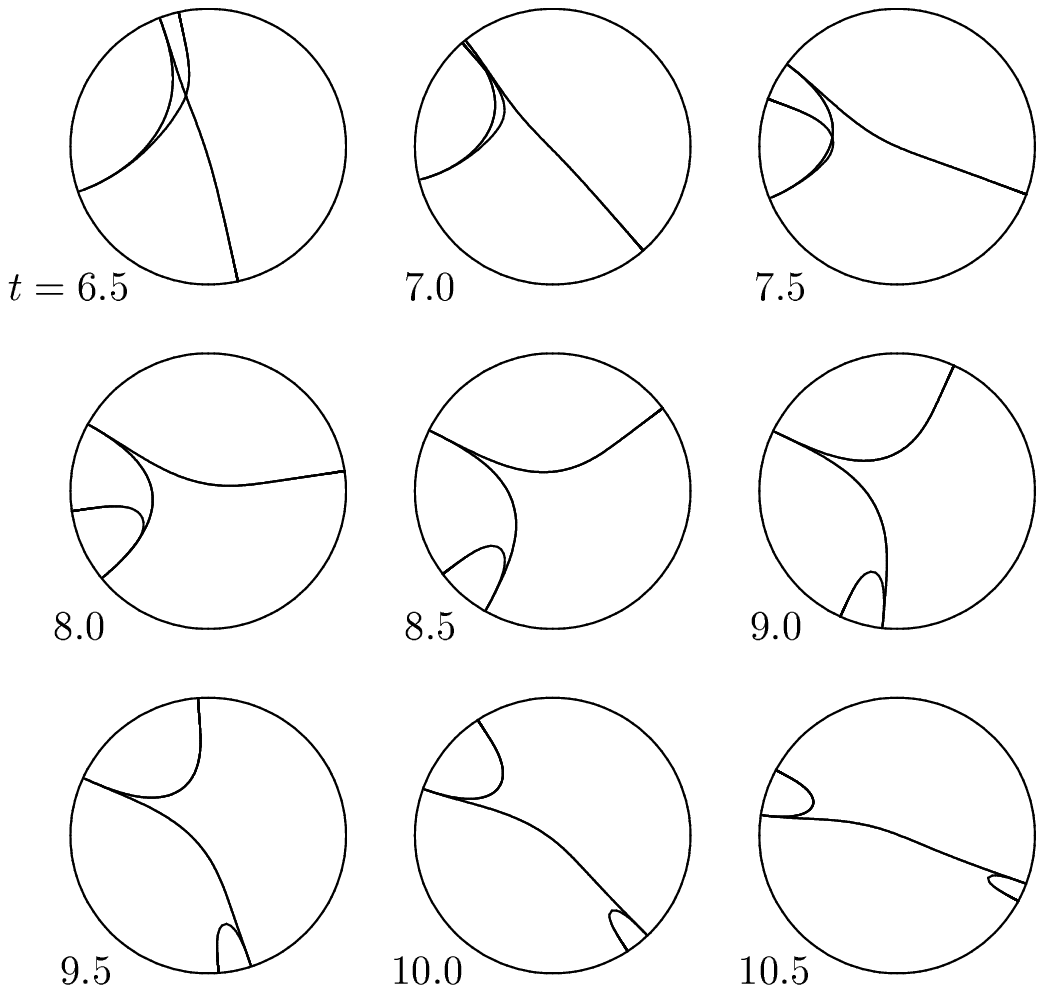,height=8cm,width=7.5cm,angle=270}
\caption{Continuation of Fig.~\ref{fig6} to larger times.} 
\label{fig7}
\end{center}
\end{figure}

\section{Summary and discussion}\label{sect5}

In this paper, we have reconsidered the simplest version of the GN model in the large $N$ limit --- the massless model with discrete chiral symmetry,
Eq.~(\ref{I1}). Following a
suggestion of Dashen, Hasslacher and Neveu we have first studied kink-antikink scattering, using the TDHF language.
A full analytical solution has uncovered an interesting interplay between the scalar potentials of kink and antikink which repell each other
and the fermion constituents which travel in one direction only, being exchanged during the collision. From a technical point of view, we have found that
unlike our starting point, the DHN breather, the scattering solution belongs to the simple class of type I solutions in which each single
particle orbit contributes a term proportional to the full self-consistent potential $S(x,t)$ to the chiral condensate. This complements earlier 
known solutions of type I, the vacuum, the kink and the kink crystal. Type I solutions are singled out by the fact that the TDHF equations
reduce to a non-linear Dirac equation. In this way we were able to make contact with earlier studies of the classical $N=1$ GN model
and take advantage of the techniques which have been developed there. We have reproduced the finding of Neveu and Papanicolaou
that the square of the self-consistent potential is related to the sinh-Gordon equation. The linearized version of the latter equation is nothing but the
Klein-Gordon equation for the $\sigma$ meson of the GN model. This supports a Skyrme-type picture of the baryon in the non-chiral GN model.
All presently known type I solutions can be identified (up to a trivial sign ambiguity) with well-known solitons of
the sinh-Gordon equation, an observation which would have saved a lot of guesswork, had it been known before. Even more interesting is perhaps the fact that
$N$ soliton solutions and the Lax representation of the associated linear problem are known explicitly for the sinh-Gordon equation. This 
should enable us to generalize the two-soliton solution of the present paper to the scattering problem of $N$ composite, relativistic bound states.
Another possible generalization of the present work would be to consider type II solutions, thereby covering all analytically known HF and TDHF
solutions of both the massless and the massive (non-chiral) GN models. The methods of Ref.~\cite{L16} are also developed for the $N=2$ classical GN model,
relevant for type II solutions of the large $N$ quantum theory. Here however, the formalism is significantly more involved, and one apparently
needs coupled, non-linear differential equations for several functions, including the scalar potential.

So far we have not yet mentioned  the chiral GN model or Nambu--Jona-Lasinio model in 2 dimensions (NJL$_2$). In the massless NJL$_2$ model, transparent
potentials appear in the context of the massless baryon \cite{L21} and the Shei bound state \cite{L26}. Generalizations to periodic, finite gap potentials include
the chiral spiral \cite{L27} and the twisted kink crystal \cite{L28}. Basar and Dunne have identified the non-linear Schr\"odinger equation as the 
relevant equation for the HF potential, using completely different techniques from the ones employed in the present paper. In the massive NJL$_2$ model,
approximate results from the derivative expansion indicate that the sine-Gordon equation is relevant near the chiral limit, but needs to be replaced by
increasingly complicated, higher order differential equations with increasing bare fermion mass \cite{L13}. Here, the baryon potentials are apparently
not transparent, so that we are dealing with neither type I nor type II solutions, nor with solutions of any finite type, for that matter.
The common theme of all of these efforts is the search for a closed, classical theory of the self-consistent
potential, the relevant saddle point in the functional integral approach. In a sense, one might identify this effort with the search for Witten's 
``master field" in the large $N$ limit of quantum chromodynamics \cite{L29} where unfortunetely little progress has been made until now.

The close relationship between the GN model and the sinh-Gordon equation gave us the clue for yet another kind of mapping. It is known that the sinh-Gordon
model and strings in AdS$_3$ are related by some kind of gauge transformation, using the Pohlmeyer reduction. In the same vein, we have constructed a gauge
transformation mapping type I GN solutions onto classical strings in AdS$_3$. Although this has nothing to do with the celebrated AdS/CFT correspondence,
it is rather intriguing. When mapping the sinh-Gordon model to string theory, one has to use artificial
spinor fields from the associated Lax pair. In our case, the spinors have a more direct physical meaning as TDHF solutions. Another difference to
the string/sinh-Gordon correspondence is a singularity in the gauge transformation at the zeros of $S$. This yields folded strings whose spikes 
must always lie on the boundary of AdS$_3$. The existence of such a mapping seems to be closely related to the symmetries
of the models. However, the fact that two models have the same symmetry is in general not sufficient to conclude that their dynamics is the same.
Here we have shown how to map a fermionic quantum field theory in the large $N$ limit onto a classical string theory in a 
fully explicit manner. Whether this is of any practical use remains to be seen. If nothing else, it provides us with a novel way of visualizing TDHF
single particle spinors, which contain more detailed information than the self-consistent potential.
\vskip 0.5cm
\section*{Acknowledgement}
\vskip 0.2cm
M. T. thanks Antal Jevicki and Kewang Jin for a clarifying correspondence about their work, and Johanna Erdmenger and Hans-J\"urgen Pirner for
helpful discussions. This work has been supported in part by the DFG under grant TH 842/1-1.

\vskip 0.5cm
{\bf Appendix: Definition and normalization of $\psi_b$}
\vskip 0.2cm 
Here we motivate the particular choice for the analytically continued TDHF spinor $\psi_b$, Eq.~(\ref{81}). Consider first a real spectral parameter 
$\zeta$. The $N$ soliton solution of the TDHF equation can be factored as
\begin{equation}
\psi = \varphi e^{i(\bar{z}\zeta-z/4\zeta)}
\label{D1}
\end{equation}
where $\varphi$ is a solution of the reduced Dirac equation
\begin{eqnarray}
-2i \varphi_{1,z} - \frac{1}{2\zeta} \varphi_1 & = &  S \varphi_2 ,
\nonumber \\
2i \varphi_{2,\bar{z}} - 2 \zeta \varphi_2 & = & S \varphi_1.
\label{D2}
\end{eqnarray}
This  merely expresses the fact that the potentials are reflectionless.  If $\varphi$ solves Eq.~(\ref{D2}) with spectral parameter $\zeta$, then
\begin{equation}
\gamma_5 \varphi^* = \left( \begin{array}{c} -\varphi_1^* \\ \varphi_2^* \end{array} \right)
\label{D3}
\end{equation}
solves the corresponding equation with spectral parameter $-\zeta$. Since we wish to introduce imaginary $\zeta$ parameters, we avoid 
complex conjugation and write the scalar condensate and the reflection property under $\zeta \to - \zeta$ in the analytic form
\begin{eqnarray}
\bar{\varphi}(\zeta) \varphi(\zeta) & = & \varphi_1(\zeta,-i)\varphi_2(\zeta,i) + \varphi_2(\zeta,-i)\varphi_1(\zeta,i),
\nonumber \\
\varphi_1(-\zeta,i) & = & - \varphi_1(\zeta,-i),
\nonumber \\
\varphi_2(-\zeta,i) & = & \varphi_2(\zeta,- i).
\label{D4}
\end{eqnarray}
We also need the normalization condition (\ref{82}) for $\psi_a$ which translates into
\begin{equation}
\bar{\varphi}(\zeta) \varphi(\zeta) = - \zeta S.
\label{D5}
\end{equation}
From Eqs.~(\ref{D4}) one deduces the relations
\begin{eqnarray}
\bar{\varphi}(-\zeta) \varphi(\zeta) & = & \bar{\varphi}(\zeta) \varphi(-\zeta) \ = \ 0,
\nonumber \\
\bar{\varphi}(-\zeta) \varphi(-\zeta) & = & - \bar{\varphi}(\zeta) \varphi(\zeta).
\label{D6}
\end{eqnarray}
In this form, all reflection properties also hold for complex $\zeta$. $\psi_b$ can be written as
\begin{equation}
\psi_b = \frac{1}{\sqrt{2}} \left[ \varphi(\zeta=-i\lambda/2) e^{\cal B} + i \varphi(\zeta=i\lambda/2)e^{-{\cal B}}\right]
\label{D7}
\end{equation}
with 
\begin{equation}
{\cal B} = \frac{\bar{z} \lambda}{2} + \frac{z}{2\lambda}.
\label{D8}
\end{equation}
Using Eqs.~(\ref{D5}) and (\ref{D6}) for imaginary $\zeta$, the condensate of $\psi_b$ becomes
\begin{eqnarray}
\bar{\psi}_b \psi_b & = & \frac{1}{2} \left[ \bar{\varphi}(i\lambda/2) \varphi(-i\lambda/2) e^{2{\cal B}}
+ \bar{\varphi}(-i\lambda/2)\varphi(i\lambda/2) e^{-2{\cal B}} \right.
\nonumber \\
& &  \left. + i  \bar{\varphi}(i\lambda/2)\varphi(i\lambda/2)
- i \bar{\varphi}(-i\lambda/2) \varphi(-i \lambda/2) \right]
\nonumber \\
& = & i \bar{\varphi}(i\lambda/2) \varphi(i\lambda/2)
\nonumber \\
& = & \frac{\lambda}{2}S,
\label{D9}
\end{eqnarray}
proving the 2nd half of Eq.~(\ref{82}).


\begin{thebibliography}{99}
\bibitem{L1}
D. J. Gross and A. Neveu, Phys. Rev. D {\bf 10}, 3235 (1974).
\bibitem{L2}
G. 't~Hooft, Nucl. Phys. B {\bf 72}, 461 (1974).
\bibitem{L3}
R. F. Dashen, B. Hasslacher, and A. Neveu, Phys. Rev. D {\bf 12}, 2443 (1975).
\bibitem{L4}
J. Feinberg, Ann. Phys. {\bf 309}, 166 (2004).
\bibitem{L5}
M. Thies and K. Urlichs, Phys. Rev. D {\bf 67}, 125015 (2003).
\bibitem{L6}
M. Thies, J. Phys. A: Math. Gen. {\bf 39}, 12707 (2006) and references therein.
\bibitem{L7}
V. Sch\"on and M. Thies, {\it At the Frontier of Particle Physics: Handbook of QCD, Boris Ioffe Festschrift},
vol. 3, ed. M. Shifman (Singapore: World Scientific), ch. 33, p. 1945 (2001).
\bibitem{L8}
P. A. M. Dirac, Proc. Camb. Philos. Soc. {\bf 26}, 376 (1930).
\bibitem{L9}
W. Brendel and M. Thies, Phys. Rev. D {\bf 81}, 085002 (2010). 
\bibitem{L9a}
E. Witten, Nucl. Phys.  B {\bf  160}, 57 (1979).
\bibitem{L10}
R. Pausch, M. Thies, and V. L. Dolman, Z. Phys. A {\bf 338}, 441 (1991).
\bibitem{L11}
M. Thies, Phys. Rev. D {\bf 69}, 067703 (2004).
\bibitem{L12}
O. Schnetz, M. Thies, and K. Urlichs, Ann. Phys. {\bf 314}, 425 (2004).
\bibitem{L13}
M. Thies and K. Urlichs, Phys. Rev. D {\bf 71}, 105008 (2005).
\bibitem{L14}
J. Feinberg and S. Hillel, Phys. Rev. D {\bf 72}, 105009 (2005).
\bibitem{L15}
O. Schnetz, M. Thies, and K. Urlichs, Ann. Phys. {\bf  321}, 2604 (2006).
\bibitem{L16}
A. Neveu and N. Papanicolaou, Commun. Math. Phys. {\bf 58}, 31 (1978).
\bibitem{L17}
A. Jevicki, K. Jin, C. Kalousios, and A. Volovich, JHEP {\bf 0803}, 032 (2008).
\bibitem{L18}
A. Jevicki and K. Jin, Int. J. Mod. Phys. A {\bf 23}, 2289 (2008).
\bibitem{L19}
A. Jevicki and K. Jin, JHEP {\bf 0906}, 064 (2009).
\bibitem{L19a}
R. Jackiw and C. Rebbi, Phys. Rev. D {\bf 13}, 3398 (1976).
\bibitem{L19b}
F. Karbstein and M. Thies, Phys. Rev. D {\bf 76}, 085009 (2007).
\bibitem{L20}
R. H. Goodman and R. Haberman, Siam J. Appl. Dyn. Syst. {\bf 4}, 1195 (2005).
\bibitem{L21}
L. L. Salcedo, S. Levit, and J. W. Negele, Nucl. Phys. B {\bf 361}, 585 (1991).
\bibitem{L22}
R. Blankenbecler and D. Boyanovsky, Phys. Rev. D {\bf 31}, 2089 (1985).
\bibitem{L23}
A. D. Polyanin and V. F. Zaitsev, {\em Handbook of nonlinear partial differential equations},
Chapman \& Hall/CRC (2004), p. 225.
\bibitem{L24}
M. J. Ablowitz, D. J. Kaup, A. C. Newell, and H. Segur, Stud. Appl. Math. {\bf 53}, 249 (1974).
\bibitem{L25}
K. Pohlmeyer, Commun. Math. Phys. {\bf 46}, 207 (1976).
\bibitem{L26}
Sun-Sheng Shei, Phys. Rev. D {\bf 14}, 535 (1976).
\bibitem{L27}
V. Sch\"on and M. Thies, Phys. Rev. D {\bf 62}, 096002 (2000).
\bibitem{L28}
G. Basar and G. V. Dunne, Phys. Rev. D {\bf 78}, 065022 (2008).
\bibitem{L29}
E. Witten, in {\em Recent developments in gauge theories}, 1979 Cargese lectures, ed. G. 't~Hooft et al.,
Plenum Press, N.Y. (1980).
\end{thebibliography}
\end{document}